Running title:

**In vitro reconstitution of actin-septin filament assembly**

Title:

**Purification of recombinant human and Drosophila septin hexamers for TIRF assays of actin-septin filament assembly**

Manos Mavrakis[1]*, Feng-Ching Tsai[2, 3] and Gijsje H. Koenderink[2]*

[1] Aix Marseille Université, CNRS, Centrale Marseille, Institut Fresnel, UMR 7249, Faculté des Sciences Saint-Jérôme, 13013 Marseille, France

[2] FOM Institute AMOLF, Science Park 104, 1098 XG Amsterdam, The Netherlands

[3] Laboratoire Physico Chimie Curie, Institut Curie, PSL Research University, CNRS UMR168, 75005, Paris, France; Sorbonne Universités, UPMC Univ Paris 06, 75005, Paris, France

*Corresponding authors:

e-mail address: manos.mavrakis@univ-amu.fr;  g.koenderink@amolf.nl

**Abstract**

Septins are guanine nucleotide binding proteins that are conserved from fungi to humans. Septins assemble into hetero-oligomeric complexes and higher-order structures with key roles in various cellular functions including cell migration and division. The mechanisms by which septins assemble and interact with other cytoskeletal elements like actin remain elusive. A powerful approach to address this question is by cell-free reconstitution of purified cytoskeletal proteins combined with fluorescence microscopy. Here, we describe procedures for the purification of recombinant Drosophila and human septin hexamers from *Escherichia coli* and reconstitution of actin-septin co-assembly. These procedures can be used to compare assembly of Drosophila and human septins and their co-assembly with the actin cytoskeleton by total internal reflection fluorescence (TIRF) microscopy.

**INTRODUCTION**

Since their discovery in budding yeast more than forty years ago (Hartwell, 1971; Hartwell, Culotti, Pringle, & Reid, 1974), septin GTP-binding proteins have been shown to be present in all eukaryotes except plants (Nishihama, Onishi, & Pringle, 2011; Pan, Malmberg, & Momany, 2007). Although several groups encountered mammalian septin genes during their studies in the early 1990s (for example, (Nakatsuru, Sudo, & Nakamura, 1994)), the identification and functional analysis of Drosophila septins (Fares, Peifer, & Pringle, 1995; Neufeld & Rubin, 1994) established that septin family proteins existed in animals and not only in budding yeast, and also provided strong evidence for roles of septins not related to cytokinesis. Soon after came the first isolation of native septin complexes from Drosophila embryonic extracts using an immunoaffinity approach (Field et al., 1996), which provided the first evidence that septin heteromeric complexes are able to polymerize into filaments in vitro. This study was followed by the isolation of endogenous heteromeric septin complexes from yeast extracts with a similar immunoaffinity protocol (Frazier et al., 1998), as well as from mammals, both by rat brain fractionation (Hsu et al., 1998) and by immunoisolation from mouse brain and HeLa cells (Kinoshita, Field, Coughlin, Straight, & Mitchison, 2002). More recently, a new protocol was reported for the purification of endogenous septin complexes from Drosophila embryonic extracts with a fractionation approach (Huijbregts, Svitin, Stinnett, Renfrow, & Chesnokov, 2009).

Understanding the requirement and the precise role of different septin proteins for the formation and stability of septin complexes, as well as for their capacity to form filaments necessitated the co-expression of septins in heterologous systems. Thus, work during the 2000s focused on the co-expression of recombinant septins in bacteria and in insect cells, using genetically encoded tags on one (in most studies) or two septins to facilitate protein purification. Budding yeast septin complexes were purified from bacteria (Bertin et al., 2008; Farkasovsky, Herter, Voss, & Wittinghofer, 2005; Garcia et al., 2011; Versele et al., 2004), mammalian septin complexes were isolated from bacteria (Sheffield et al., 2003; Sirajuddin et al., 2007) and insect cells (Kinoshita et al., 2002), C. elegans septin complexes from bacteria and insect cells (John et al., 2007), and Drosophila septin complexes were also expressed in and purified from bacteria (Huijbregts et al., 2009; Mavrakis et al., 2014) and insect cells (Huijbregts et al., 2009). Biochemical analysis and electron microscopy of the purified recombinant complexes in these studies, as well as of mammalian septin complexes immuno-purified (Sellin, Sandblad, Stenmark, & Gullberg, 2011) or affinity-purified (Kim, Froese, Estey, & Trimble, 2011) from mammalian cell cultures, together with the crystal structure of a human septin complex (Sirajuddin et al., 2007), have altogether established that septins from all organisms form rod-shaped complexes containing two, three or four septins with each present in two copies, forming a tetramer (C.elegans), hexamer (Drosophila and human) or octamer (budding yeast and human), respectively.

The combination of electron microscopy (EM) with in vitro reconstitution of septin filament assembly in low-salt conditions (<100 mM KCl) using recombinant purified complexes has been instrumental and is still one of the most powerful approaches for deciphering how septin protomers assemble into filaments and how filaments organize into higher-order structures. The main drawback of EM approaches is that they provide snapshots of septin organization and do not allow studies on how freely-diffusing septin protomers dynamically assemble, grow and organize into filament assemblies. Fluorescence-based assays using purified recombinant septin complexes present great potential in this aspect since they can combine molecular specificity (when tagging specific septin subunits) with a wide range of fluorescence-based techniques that enable studies across different spatial and temporal scales.

At present there is only a handful of reports using fluorescently tagged recombinant septin complexes for studying septin assembly, almost exclusively for budding yeast septins. Three approaches are being used for fluorescent tagging: (1) genetic fusion of a specific septin subunit with a fluorescent protein, such as mEGFP or mCherry (Bridges et al., 2014; Sadian et al., 2013), (2) genetic fusion of a specific septin subunit with a SNAP-tag and its further derivatization with fluorescent dyes (Renz, Johnsson, & Gronemeyer, 2013), or (3) chemical conjugation at lysines or cysteines of purified complexes with Alexa Fluor dyes (Booth, Vane, Dovala, & Thorner, 2015; Mavrakis et al., 2014). The recent development of a TIRF imaging assay using purified GFP-tagged yeast septin octamers on supported lipid bilayers (Bridges et al., 2014) provided the first fluorescence microscopy-based quantitative assay for studying the kinetics of yeast septin filament assembly, highlighting the promise of this approach.

The mechanisms that control septin assembly into complexes and higher-order filamentous structures and the regulation of septin structures and their dynamics are still largely unclear. Especially little in vitro work has been done on human and Drosophila septins, so it remains unknown how animal septin assembly differs from budding yeast septin assembly. In addition, septin interactions with other cytoskeletal components remain elusive. Here, we describe procedures to purify recombinant Drosophila and human septin hexamers from Escherichia coli. We describe procedures for fluorescent tagging (using either genetic GFP fusions or chemical conjugation with organic dyes), and provide protocols for in vitro reconstitution of actin and septin assembly in surface assays amenable to high-resolution imaging by TIRF microscopy.

## 1. Cloning strategy for recombinant septin complex production in bacteria

To isolate stoichiometric hexameric septin complexes we combine the extended pET-MCN (**pET M**ulti-**C**loning and expressio**N**) series as a septin co-expression system (Diebold, Fribourg, Koch, Metzger, & Romier, 2011) with a two-tag purification scheme. To this end, we use two vectors, pnEA-vH and pnCS. pnEA-vH harbors the central subunit of the hexamer (DSep1 for Drosophila septins or hSep2 for human septins) under the control of the T7 promoter, with a TEV-cleavable 6xHis-tag fused to its N-terminus. pnCS harbors the other two septin genes (DSep2 and Peanut for Drosophila septins, or hSep6 and hSep7 for human septins) under the control of a single T7 promoter (Figure 1). Using appropriate PCR primers we fuse the C-terminus of the terminal subunit of the hexamer (Pnut for Drosophila septins or hSep7 for human septins) to a noncleavable eight-amino-acid Strep-tag II (WSHPQFEK, 1058 Da).

To generate the pnCS vector harboring two septin genes under the control of a single promoter, we use bidirectional cloning to concatenate two pnCS vectors encoding one septin gene each. Each septin gene in the pnCS vector is flanked by SpeI on the 5' side of the MCS and by XbaI on the 3' side. Using standard cloning procedures (Green & Sambrook, 2012) we double-digest the donor plasmid with SpeI / XbaI, ligate the insert to SpeI-digested acceptor plasmid and select the clones with the correct insert orientation using restriction analysis (Figure 1).

We have used this strategy for successful coexpression of both Drosophila and human septin hexamers. The flexibility of the pET-MCN series (detailed in (Diebold et al., 2011)) enables rapid screening to test how the position of the tags, the order of septin genes, or their expression under a single or multiple promoters affects the quantity and stoichiometry of the resulting complex.

[Insert Figure 1 here]

**Figure 1**. Overview of cloning strategy for co-expressing recombinant septins using the pnEA-vH and pnCS vectors of the pET-MCN series (Diebold et al., 2011). Here this strategy is used for the generation of recombinant Drosophila (DSep1-DSep2-Pnut) and human (hSep2-hSep6-hSep7) septin hexamers.

## 2. Expression of recombinant septin complexes in bacteria

We co-express $His_6$-hSep2, hSep6 and hSep7-Strep for producing human septin hexamers (hSep2-hSep6-hSep7), and $His_6$-DSep1, DSep2 and Pnut-Strep for producing Drosophila septin hexamers (DSep1-DSep2-Pnut).

**Materials and reagents**

Solutions are prepared in water unless stated otherwise.

- plasmids for co-expression of septins:

Drosophila septins: His$_6$-DSep1 in pnEA-vH, DSep2/Pnut-Strep in pnCS

Drosophila GFP-labeled septins: His$_6$-DSep1 in pnEA-vH, mEGFP-DSep2/Pnut-Strep in pnCS

human septins: His$_6$-hSep2 in pnEA-vH, hSep6/hSep7-Strep in pnCS

- E.coli BL21(DE3) competent cells (200131, Agilent Technologies), -80°C
- Ampicillin (A0166, Sigma), 100 mg/mL, -20°C
- Spectinomycin (S4014, Sigma), 100 mg/mL, -20°C
- LB medium (L3022, Sigma), RT
- LB agar (L2897, Sigma), RT
- LB agar plates containing both ampicillin and spectinomycin at 100 μg/mL each, 4°C
- SOC medium (S1797, Sigma), -20°C
- Terrific Broth (091012017, MP Biomedicals), RT
- IPTG (EU0008-C, Euromedex), 1 M, -20°C
- 2L Erlenmeyer flasks

**Day 0. Heat-shock transformation of bacteria**

We use standard procedures (Green & Sambrook, 2012) to co-transform E.coli BL21(DE3) competent cells with the two plasmids encoding all three septin genes, and grow colonies on LB agar plates at 37°C overnight.

**Day 1. Bacterial pre-culture**

1. Using a sterile pipette tip, select a single colony from your LB agar plate. Drop the tip into liquid LB containing both antibiotics (1000x dilution of the antibiotic stock solutions) and swirl. Calculate the volume of the pre-culture given that you will use 1/50 of the total volume of the culture for inoculation.

2. Incubate the bacterial culture at 37°C for 12-16 h in a shaking incubator to prepare your pre-culture. For long-term storage of the co-transformed bacteria, mix your pre-culture with glycerol to make a 50% v/v glycerol stock and store it at -80°C.

**Day 2. Bacterial culture for producing dark (unlabeled) septin complexes**

The C-termini of both Peanut (when co-expressed with DSep1 and DSep2) and of hSep7 (when co-expressed with hSep2 and hSep6) are prone to degradation by bacterial proteases during protein expression (Mavrakis et al., 2014). The presence of the Strep-tag II helps isolate complexes containing full-length Pnut/hSep7. To minimize degradation, we developed a protocol whereby each bacterial cell produces less protein (short induction time) but with less degradation. We compensate for the smaller protein yield per cell by using Terrific Broth to grow bacteria at 37°C to a high density before induction. We typically prepare 6-8 L of culture.

1. Add 700 mL Terrific Broth containing both antibiotics (1000x dilution of the antibiotic stock solutions) into each 2L Erlenmeyer flask.

2. Inoculate each flask with your pre-culture (1/50 of the culture volume).

3. Incubate at 37°C in a shaking incubator (220rpm) until the OD600 reaches 2-3 (this should take 3h-4h30).

4. Induce protein expression by adding IPTG to 0.5 mM final. Allow cells to grow for 3 h at 37°C.

5. Collect cells by centrifuging at 2,800 g and 4°C for 10 min. Pool bacterial pellets in a 50 mL Falcon tube and store at -80°C. If you want to proceed directly with protein purification, store the tube at -80°C for at least 30 min, then continue with cell lysis.

**Days 2-3. Bacterial culture for producing GFP-labeled septin complexes**

To minimize protein degradation and also allow GFP to fold, we grow cells at 37°C to a low density and then induce expression at 17°C overnight. We typically prepare 8-10 L of culture. We minimize exposure to light by covering the incubator with aluminum foil and keeping the lights off.

1. Inoculate flasks with your pre-culture as described for dark septin production.

3. Incubate at 37°C in a shaking incubator (220rpm) until the OD600 reaches 0.6-0.8 (this takes 2h30-3h).

4. Induce protein expression by adding IPTG to 0.5 mM final. Allow cells to grow overnight (16 h) at 17°C.

5. Collect and store bacterial pellets as described for dark septin production. The pellets should be yellow-greenish confirming the presence of GFP.

# 3. Purification and characterization of recombinant septin complexes from bacteria

We use a two-tag purification scheme (His$_6$-tag on the central subunit i.e. DSep1/hSep2 and a Strep-tag II on the terminal subunit i.e. Pnut/hSep7) in order to select for complexes with full-length Pnut/hSep7 and to minimize isolation of substoichiometric septin complexes (Figure 2). The nickel affinity column isolates His$_6$-DSep1/hSep2-containing complexes, whereas the Strep-Tactin (engineered streptavidin) affinity column further isolates those complexes that also contain Pnut/hSep7-Strep thus heterohexamers. A final gel-filtration step helps remove aggregates and isolate hexamers. We use the same protocol for purifying human septin hexamers (hSep2-hSep6-hSep7) and Drosophila septin hexamers (DSep1-DSep2-Pnut). When purifying mEGFP-tagged septins, we minimize exposure to light by covering columns with aluminum foil and keeping lights off.

[Insert Figure 2 here]

**Figure 2.** Schematic overview of the two-tag purification scheme for isolating stoichiometric human septin hexamers (the same applies to Drosophila septins by analogy). See text for details. Understanding the principles of septin complex assembly is an intense topic of investigation, and the presence and stability of intermediate complexes has not been fully documented. We show selected populations of human septin complexes in the cell lysate (monomers, homo- and heterodimers, heterotrimers and hexamers) based on the isolation and characterization of such recombinant and native complexes (Kim, Froese, Xie, & Trimble, 2012; Mavrakis et al., 2014; Sellin et al., 2011; Serrao et al., 2011; Sheffield et al., 2003; Sirajuddin et al., 2007; Sirajuddin, Farkasovsky, Zent, & Wittinghofer, 2009; Zent, Vetter, & Wittinghofer, 2011; Zent & Wittinghofer, 2014).

**Materials and reagents**

Solutions are prepared in water unless stated otherwise.

For the lysis buffer:

- Tris-HCl pH 8, 1 M, RT

- KCl, 4 M, RT

- OmniPur® Imidazole (5710-OP, MERCK Millipore), 1M, RT

- $MgCl_2$ 1 M, RT

- PMSF (78830, Sigma), 100 mM in ethanol, -80°C, dilute into lysis buffer immediately before use

- Lysozyme (5934-D, Euromedex), 50 mg/mL, -20°C

- cOmplete™ Protease Inhibitor Cocktail Tablets (11697498001, Roche), 4°C: use 1 tablet for 50 mL of lysis buffer

- MgSO$_4$, 2M, RT

- DNase I (10104159001, Roche), 2 g/L, -20°C

For protein purification:

- d-Desthiobiotin (D1411, Sigma), powder, 4°C: add to 2.5 mM final in StrepTrap elution buffer immediately before use

- DTT, 1 M, -20°C: add to 5 mM final in gel filtration buffer immediately before use

- HisTrap FF crude 5-mL column (17-5286-01, GE Healthcare)

- StrepTrap HP 1-mL column (28-9075-46, GE Healthcare)

- HiLoad 16/600 Superdex 200 pg column (28-9893-35, GE Healthcare)

- ÄKTA FPLC system

For protein concentration:

- Amicon Ultra 4 mL Centrifugal Filter Units with membrane NMWL of 30 kDa (UFC803024, Millipore)

- Triton X-100 5% v/v, 4°C

**Day 1. Septin complex purification**

Filter all buffers through 0.22-μm filters. Perform all purification steps at 4°C. Keep aliquots from the elution peaks after each purification step for SDS-PAGE/Coomassie staining and Western blots for monitoring protein integrity and enrichment of septin complexes. Measure the protein concentration after each purification step using absorbance measurements at 280 nm for monitoring protein loss during the purification process.

1. Resuspend the bacterial pellet in ice-cold lysis buffer (50 mM Tris-HCl pH 8, 500 mM KCl, 10 mM imidazole pH 8, 5 mM $MgCl_2$, 0.25 mg/mL lysozyme, 1 mM PMSF, cOmplete™ protease inhibitor cocktail, 0.01 g/L DNase I, 20 mM $MgSO_4$). We use 100 mL of lysis buffer for a bacterial pellet from a 6 L culture. Lyse cells on ice with a tip sonicator using 5 cycles of 30 s "ON", 30 s "OFF" (30% amplitude).

2. Clarify the lysate by centrifugation at 20,000 g for 30 min at 4°C.

3. Load the supernatant on a HisTrap FF crude column equilibrated with 5 column volumes (CV) of 50 mM Tris-HCl pH 8, 500 mM KCl, **10 mM imidazole** pH 8, 5 mM $MgCl_2$. Wash with 7CV of the same buffer, then wash with 7CV of 50 mM Tris-HCl pH 8, 500 mM KCl, **20 mM imidazole** pH 8, 5 mM $MgCl_2$ to remove nonspecifically bound untagged proteins.

4. Elute $His_6$-DSep1/hSep2-containing complexes with 7CV of 50 mM Tris-HCl pH 8, 500 mM KCl, **250 mM imidazole** pH 8, 5 mM $MgCl_2$. Collect 1-mL fractions. Pool all fractions contained in the elution peak (typically 10-15 mL).

5. Load the pooled fractions to a StrepTrap HP column equilibrated with 5CV of 50 mM Tris-HCl pH 8, 300 mM KCl, 5 mM $MgCl_2$. We use two StrepTrap HP columns in tandem. Wash with 7CV of the same buffer.

6. Elute Pnut/hSep7-Strep-containing complexes with 7CV of 50 mM Tris-HCl pH 8, 300 mM KCl, 5 mM $MgCl_2$, **2.5 mM desthiobiotin**. Collect 1-mL fractions. Pool all fractions contained in the elution peak (typically 5 mL).

7. Load the pooled fractions to a Superdex 200 HiLoad 16/60 column equilibrated with 1.2 CV of **50 mM Tris-HCl pH 8, 300 mM KCl, 5 mM $MgCl_2$, 5 mM DTT**. Elute with 1.2 CV of the same buffer. Elute for 0.25CV before collecting 1-mL fractions. Pool the fractions corresponding to the elution peak (typically 10 mL) and concentrate using passivated Amicon concentrators (see below). Measure the concentration using the elution buffer as a blank, prepare 10- or 20-µL aliquots, flash-freeze purified septin complexes in liquid nitrogen and store at −80°C. This purification protocol typically yields 1-2 mg of stoichiometric septin hexamers, which we concentrate to 10-15 µM (about 3-5 mg/mL).

We calculate septin complex concentration from absorbance measurements at 280 nm. We compute extinction coefficients from the amino acid sequences using ExPASy at http://web.expasy.org/protparam/, assuming two copies of each full-length septin (tags included) per hexamer.

*dark Drosophila septins (His$_6$-DSep1/DSep2/Pnut-Strep):*

306.9 kDa, 1 g/L = 3.3 µM, $\varepsilon$=0.545 L.g$^{-1}$.cm$^{-1}$ at 280 nm (assuming all Cys reduced)

*mEGFP-tagged Drosophila septins (His$_6$-DSep1/mEGFP-DSep2/Pnut-Strep):*

361.6 kDa, 1 g/L = 2.8 µM, $\varepsilon$=0.584 L.g$^{-1}$.cm$^{-1}$ at 280 nm (assuming all Cys reduced)

*dark human septins (His$_6$-hSEP2/hSEP6/hSEP7-Strep):*

285.7 kDa, 1 g/L = 3.5 µM, $\varepsilon$=0.565 L.g$^{-1}$.cm$^{-1}$ at 280 nm (assuming all Cys reduced)

**Day 2a. Concentration of purified septin complexes**

Septins tend to be sticky and adsorption to the membrane of the concentrator leads to significant yield loss. To improve recovery of septins during the concentration step, we passivate the concentrators by treating them with 5% v/v Triton X-100.

1. Wash the concentrator by filling with water and spinning the liquid through at 4,500 g for 10 min. Remove residual water thoroughly by pipetting.

2. Fill the concentrator with 5% v/v Triton X-100. Incubate for 2 h at RT.

3. Remove the Triton X-100 solution. Rinse the device 3 or 4 times thoroughly with water and finally spin through for 5 min at 4,500 g. The passivated device is now ready to use.

4. Add gel filtration elution buffer and spin through for 5 min at 4,500 g to equilibrate the concentrator membrane.

5. Add the gel filtration eluate and spin through until reaching the desired volume/concentration. Do cycles of spinning of 20 min at 4,500 g. Between two cycles, check the volume of the protein solution, refill with the remaining solution, mix the protein solution very gently by pipetting up and down and remove the flow through. Keep an aliquot for SDS-PAGE.

**Day 2b. Characterization of purified septin complexes**

We analyze septin prep purity and protein integrity by 12% SDS-PAGE and Western blot. For Drosophila septins, we use mouse 4C9H4 anti-Pnut (1:100, Developmental Studies Hybridoma Bank), rat anti-DSep1-95 (1:500) (Mavrakis et al., 2014) and guinea pig anti-DSep2-92 (1:500) (Mavrakis et al., 2014). For human septins, we use goat anti-hSep2 (1:500, Santa Cruz Biotechnology, sc-20408), rabbit anti-hSep6 (1:500, Santa Cruz Biotechnology, sc-20180) and rabbit anti-hSep7 (1:200, Santa Cruz Biotechnology, sc-20620). For both Drosophila and human septins, we use HRP-conjugated anti-Penta-His (1:10,000, Qiagen) and HRP-conjugated anti-Strep-tag (1:10,000, AbD Serotec).

hSep2 antibodies recognize the N-terminus of hSep2, whereas hSep6 and hSep7 antibodies recognize the C-terminus of hSep6 and hSep7, respectively. Pnut antibodies recognize the N-terminus of Peanut (Mavrakis et al., 2014), whereas DSep1-95 and DSep2-92 antibodies recognize the N-terminus of DSep1 and DSep2, respectively (Mavrakis et al., 2014). We combine these antibodies with the His- and Strep-tag antibodies that recognize N- and C-termini, respectively, to examine the integrity of each septin in the prep by Western blot.

The purity of the septin preps can be further tested by mass spectrometry both in solution and from protein bands excised from the gels. We found the N-terminal methionines of all three Drosophila septins cleaved in the bacterial preps. The only protein we identified in our preps besides septins was the bacterial chaperone DnaK which runs at 70 kDa in SDS-PAGE (Mavrakis et al., 2014).

Finally, we evaluate the quality of each septin preparation in terms of oligomer composition and hexamer content by transmission electron microscopy combined with 2D single particle image analysis (Figure 3), as described in the chapter by Aurelie Bertin.

## 4. Labeling septins for TIRF imaging

### 4a. Generating septin-GFP fusions

For TIRF imaging of fluorescent Drosophila septins, fuse mEGFP to the N-terminus of DSep2 and generate a pnCS vector harboring both mEGFP-DSep2 and Pnut-Strep, as detailed above. Co-express $His_6$-DSep1 with mEGFP-DSep2/Pnut-Strep to produce and purify fluorescent septin hexamers, $His_6$-DSep1/mEGFP-DSep2/Pnut-Strep, as detailed above.

[Insert Figure 3 here]

**Figure 3.** Characterization of purified septin complexes by SDS-PAGE (unlabeled and GFP-labeled Drosophila septin hexamers, left and right panels in A, respectively) and by 2D single particle analysis of electron microscopy images (B, see chapter by Aurelie Bertin).

**4b. Chemical labeling of purified septin complexes with Alexa Fluor dyes**

Septins can be labeled at primary amines (ε-amino group of lysines and N-terminus) using Alexa Fluor 488 succinimidyl esters. We isolate labeled septins in filamentous form, to ensure that hexamers are still polymerization-competent after labeling. Different septin complexes and septins from different species can polymerize to different extents to higher-order structures of different sizes, which might require optimization of the incubation and centrifugation times detailed below.

**Materials and reagents**

- purified septins in 50 mM Tris-HCl pH 8, 300 mM KCl, 5 mM $MgCl_2$, 5 mM DTT (septin buffer)
- Alexa Fluor 488 carboxylic acid, succinimidyl ester (NHS ester) (A20000, Invitrogen), -20°C, protected from light
- Tris-HCl pH 8, 1M, RT
- Hepes-KOH pH 8, 1M, 4°C, stored in the dark
- PD-10 columns (GE Healthcare)

**Day 1. Reaction with NHS ester and pelleting of polymerization-competent septins**

1. To exclude reactivity of NHS esters with amine and thiol groups of compounds in the septin buffer (Tris and DTT), dialyze septins into 50 mM Hepes-KOH pH 8, 300 mM KCl, 5 mM MgCl$_2$.

2. Prepare a 20 mM stock of the NHS ester in DMSO immediately before starting the reaction. Add to dialyzed septins at 4:1 molar ratio (dye:septins), mix gently and incubate for 1 h at RT in the dark. Quench the reaction by adding Tris to 20 mM Tris-HCl pH 8 final.

3. Add 50 mM Tris-HCl pH 8, 5 mM MgCl$_2$, 5 mM DTT to dilute KCl to 50 mM final and polymerize septins for 1 h at room temperature. To pellet polymerization-competent Alexa-tagged septins, centrifuge at 100,000 g for 3 h at 4°C (we use a TLA100.3 rotor with 1.5-mL tubes). The resulting pellet will be yellow-greenish confirming the reaction with Alexa Fluor 488 (AF488) NHS esters.

4. Carefully remove the supernatant and add 0.5 mL ice-cold septin buffer to the AF488-septin pellet. Allow the pellet to resuspend on ice overnight in the dark. Mix very gently the next morning by pipetting up and down.

**Day 2. Separation of AF488-septins from unreacted AF488 with a PD-10 column**

1. Pre-cool and equilibrate a PD-10 column with 25 mL septin buffer and discard the eluate.

2. Add 0.5 mL of the resuspended AF488-septins. Once the solution has entered the column, add 2.0 mL septin buffer. Discard the eluate.

3. Elute with 3.5 mL septin buffer in 0.5-mL fractions. AF488-septins elute in fractions 2-4 (1.5 mL in total). Measure the protein concentration and calculate the degree of labeling. Use the values provided by the manufacturer for the dye ($\lambda_{max}$, $\varepsilon$, correction factor at 280 nm) to correct for the contribution of the dye to the absorbance at 280 nm. Prepare 10- or 20-μL aliquots, flash-freeze AF488-septins in liquid nitrogen and store at −80°C in the dark.

## 5. Building flow cells for TIRF microscopy

To visualize actin filaments and septins by total internal reflection fluorescence (TIRF) microscopy, we design flow cells having 6 rectangular shaped channels with approximate dimensions of 22 mm in length, 2.5 mm in width and 0.1 mm in height (Figure 4). Approximately 10 μL of a protein solution can be loaded into each channel.

[Insert Figure 4 here]

**Figure 4.** A flow cell assembled by melting strips of Parafilm between a cleaned and passivated coverslip and a glass slide.

**Materials and reagents**

● $H_2O_2$ (35% in water, 95299, Sigma), 4°C

● $NH_4OH$ (30% in water, 320145, Sigma), RT

● Dichlorodimethylsilane (DDS), (40140, Sigma), RT

● Trichloroethylene (TCE), (251402, Sigma), RT

● Pluronic® F-127 (P2443, Sigma), 10% in DMSO, RT

To completely dissolve Pluronic® F-127 powder in DMSO, we warm up the solution to 45°C. For flow cell treatment, we dilute to 1% (w/v) in F-buffer right before use.

● Parafilm (PM996, Bemis)

● Microscope glass slide (24 × 60 mm, thickness #1, 0.15mm, Menzel Gläser)

● Microscope coverslips (22 × 40 mm, thickness #1, 0.15mm, Menzel Gläser)

**5a. Cleaning and silanizing microscope glass slides/coverslips**

To remove organic residues from the glass substrates of the flow cells, we clean the microscope slides and coverslips in a base-piranha solution. Additionally, the piranha treatment generates free hydroxyl groups on the glass substrates, making them hydrophilic. To prevent nonspecific protein adhesion to the surfaces, we coat them with a hydrophobic layer of dimethylsilane.

1. Load microscope glass slides and coverslips into a Teflon rack.

2. Place the Teflon rack into a glass beaker filled with Milli-Q water. Rinse the glass slides and coverslips twice, 5 min each, with Milli-Q water in a bath sonicator.

3. Transfer the Teflon rack to a new glass beaker. The glass beaker should be large enough to cover the glass slides and coverslips with the base-piranha solution (Milli-Q water, 30% $NH_4OH$ and 35% $H_2O_2$ at a 5:1:1 volume ratio). In a fume hood, fill the glass beaker with five parts Milli-Q water, heat on a hot plate to 80°C. Add one part $NH_4OH$ and then slowly add one part $H_2O_2$. Gently mix the solution by moving the Teflon rack up-and-down in the beaker. Once the piranha solution reaches a temperature of 60°C, bubbles form in the solution, indicating an active reaction. Allow the solution to react for 30 min.

*Tip*: The base-piranha solution is a dangerous reaction, thus wear gloves and safety goggles, and avoid spills.

*Tip*: Make fresh base-piranha solution for each use because the solution decomposes. The used solution can be left in the hood overnight. Once the reaction has completed, only water is left, which can be safely discarded down the sink.

4. Rinse the slides and coverslips as in step 2.

5. To activate the hydroxyl groups on glass substrates for the silanization reaction, transfer the slides and coverslips to a glass beaker filled with a 0.1M KOH solution and sonicate for 15 min in a bath sonicator.

*Note*: If need be, the substrates can be stored in the 0.1M KOH solution for up to one month, otherwise we proceed to the following steps.

6. Transfer the slides and coverslips to a glass beaker filled with Milli-Q water, sonicate for 5 min, blow-dry completely with nitrogen gas, and silanize immediately.

7. In a fume hood, prepare a solution of 0.05% v/v DDS in TCE in a glass beaker that is large enough to cover the slides and coverslips with the solution.

*Note*: Both DDS and TCE are volatile and toxic, thus wear gloves, do not inhale and work always in the fume hood.

8. Transfer the slides and coverslips into the glass beaker filled with the DDS/TCE solution and incubate for 1 h, without stirring.

*Note*: Carefully dispose used DDS/TCE solution according to local laboratory safety regulations.

9. Transfer the slides and coverslips in a glass beaker filled with methanol and sonicate for 15 min in a bath sonicator.

10. Blow-dry completely the slides and coverslips with nitrogen gas and store in a clean sealed container for up to one week.

**5b. Constructing flow cells**

Here we describe how to construct a flow cell by melting strips of Parafilm between a glass slide and a coverslip.

1. Blow-dry a silanized glass slide with nitrogen gas.

2. Cut a piece of Parafilm that is large enough to cover the glass slide. Place the Parafilm layer on the glass slide and press firmly to ensure the layer adheres to the glass slide. The thickness of the Parafilm layer sets the height of the flow channels.

3. Use a scalpel to cut the Parafilm layer along the edges of the glass slide, trimming away excess Parafilm along the edges.

4. Use a scalpel to cut 2.5 mm wide strips of the Parafilm layer along the short edge of the glass slide. Cut 11 strips, centered with respect to the glass slide. Cut firmly to ensure the edges of the adjacent strips are well separated.

5. Remove every second strip by using a forceps. The width of the removed strips sets the width of the flow channels.

6. Place a silanized glass coverslip on top of the strips, centered with respect to the glass slide.

7. Place the assembly on a hot plate set at 120°C. Once the Parafilm strips are melted, use a forceps and gently press the coverslip to maximize the contact area between the strips and the coverslip. Do not apply too much pressure that can break the coverslip. Remove the assembly from the hot plate and leave it to cool at room temperature.

*Tip*: While melting the assembly, air bubbles can appear at the contact surface between the Parafilm strips and the glass coverslip. Apply gentle pressure on the coverslip with a forceps to remove the air bubbles, which could cause leakage between adjacent channels.

8. Flow 1% Pluronic® F-127 into channels, incubate for 5 min and rinse with the actin polymerization buffer (F-buffer) (see section 6b for the composition of F-buffer). The flow cell is ready to use.

*Note*: Silanized glass substrates are hydrophobic, which hampers inflow of sample solutions. We thus treat channels with Pluronic F-127 to render the surface hydrophilic enough to be able to flow in the aqueous solution.

*Note:* Pluronic® F-127 is a copolymer composed of a hydrophobic block of polypropylene glycol flanked by two hydrophilic blocks of polyethylene glycol, thus having surfactant properties.

## 6. *In vitro* reconstitution of actin-septin filament assembly

**Materials and reagents**

Solutions are prepared in water unless stated otherwise.

● DTT (D0632, Sigma), 1M, -20°C

● MgATP, 100 mM, -80°C

*Note*: To obtain MgATP, prepare a 200 mM solution of Na$_2$ATP (ATP disodium salt hydrate, A2383, Sigma) by dissolving Na$_2$ATP powder in Milli-Q water and adjusting the pH of the solution to pH 7.4 using NaOH solution. Mix the Na$_2$ATP solution 1:1 with a 200 mM solution of MgCl$_2$ to obtain 100 mM MgATP.

● Methylcellulose (M0512, Sigma), 1% (w/v), RT

● Protocatechuate acid (PCA) (03930590-50MG, Sigma), 100 mM in water and adjusted to pH 9 using NaOH, -80°C

● Protocatechuate 3,4-dioxygenase (PCD) (P8279-25UN, Sigma), 5 µM in 50% v/v glycerol, 50 mM KCl, 100 mM Tris-HCl pH 8, -80°C

● Trolox (238813-1G, Sigma), 100 mM, -20°C

   To prepare a 100 mM Trolox solution:

   1. Dissolve 0.1 g of Trolox powder in 430 µL methanol

2. Add 3.2 mL of Milli-Q water

3. Add 360 μL of 1 M NaOH (The solution turns yellowish)

- VALAP (a mixture of vaseline, lanolin and paraffin with equal weight), RT

### 6a. Preparing protein solutions

G-actin is purified from rabbit skeletal muscle by a standard procedure including a final gel filtration step on a HiPrep 26/60 Sephacryl S-200 HR column (GE Healthcare) (Pardee & Spudich, 1982). G-actin is stored at -80°C in G-buffer (2mM Tris-HCl, pH 7.8, 0.2 mM Na$_2$ATP, 0.2 mM CaCl$_2$, and 2 mM DTT). G-actin is fluorescently labeled with Alexa Fluor® 594 carboxylic acid, succinimidyl ester (AF594-G-actin) (Soares e Silva et al., 2011) and is also stored at -80°C in G-buffer. Dark and fluorescent septins are stored in septin buffer as detailed above. Before use, aliquots of actin and septins are thawed on ice and cleared for 5 min at 120,000 g at room temperature in a Beckman airfuge. Finally, protein concentrations are determined by measuring the absorbance of the protein solutions at 280 nm, using an extinction coefficient of 1.1 L.g$^{-1}$.cm$^{-1}$ for G-actin (1 g/L = 23.8 μM). Proteins are kept on ice and used within one week.

### 6b. Buffers and components

The final assay buffer (referred to as F-buffer, or actin polymerization buffer) contains the following components:

- 20 mM imidazole-HCl, pH 7.4
- 50 mM KCl
- 2 mM MgCl$_2$
- 0.1 mM MgATP

- 1 mM DTT
- 1 mM Trolox
- 2 mM PCA
- 0.1 μM PCD
- 0.1% (w/v) methylcellulose

*Note*: When mixing proteins with the above components to reach the composition of the F-buffer, take into consideration that the septin buffer contains 300 mM KCl and 5 mM $MgCl_2$.

*Note*: Trolox is included in the solution to quench triplet states and thus to prevent photobleaching due to the reactions of the triplet states with oxygen free radicals (Cordes, Vogelsang, & Tinnefeld, 2009). PCA-PCD is a substrate-enzyme pair that scavenges oxygen free radicals and thus minimizes photobleaching (Shi, Lim, & Ha, 2010).

*Note*: For TIRF imaging, we employ a 0.1% (w/v) solution of methylcellulose, which exerts an entropic force pushing actin-septin filaments towards the surface to facilitate observation by TIRF imaging. However, we recommend to use at most 0.2% (w/v) methylcellulose, since higher methylcellulose concentrations induce the formation of actin bundles (Popp, Yamamoto, Iwasa, & Maeda, 2006).

**6c. Reconstituting actin-septin filament assembly**

Here we describe a procedure to prepare samples in which actin and septins co-polymerize. To visualize actin and septins, we dope actin with AF594-G-actin (10% molar label ratio) and septins with AF488-septins or GFP-septins (10% molar label ratio) (Figure 5). We prepare protein samples with a final volume of 10 μl. We provide as an example the sample preparation for co-polymerizing 1 μM actin with 1 μM septins.

1. Prepare a "master buffer" containing 5-fold higher concentrations of all the components of F-buffer apart from PCD, taking into account the contribution of KCl and MgCl$_2$ from the septin solution that will be added into the final mixture to reach the desired septin concentration.

2. Mix labeled and unlabeled G-actin to a final concentration of 5 μM in G-buffer with a 10% molar label ratio.

*Note*: Given that dense G-actin solutions (>= 5 mg/ml) are quite viscous and difficult to mix, we dilute G-actin to an intermediate concentration.

3. Mix labeled and unlabeled septins to a final concentration of 6 μM in septin buffer with a 10% molar label ratio.

4. In one Eppendorf tube, add 4.1 μL of Milli-Q water, 2 μL of master buffer (5x), 0.2 μL of PCD and 1.7 μL of labeled septins and mix well.

*Tip*: We typically dilute the septin solution 6-fold into the final mixture in order to obtain 50 mM KCl from the septin buffer. When performing a series of samples in which actin concentration stays constant but septin concentration varies, one can prepare septin solutions having 6-fold higher concentrations than the final desired concentrations by diluting septins in septin buffer, and thus pipette 1.7 μL of each septin dilution in every sample.

5. In another Eppendorf tube, place 2 μL of labeled actin.

6. Load the septin mixture from step 4 into the G-actin containing tube, mix the two solutions thoroughly by aspirating up and down 3 times and immediately load the mixed solution into one flow channel.

*Note*: Given that G-actin polymerizes into F-actin immediately when mixed with salts, it is important to perform the above step quickly.

7. Seal the two open ends of the channels with VALAP using a cotton-tipped applicator.

*Note*: Before use, melt VALAP at temperatures exceeding 80°C. We typically keep a small beaker with liquid VALAP on a hot plate (120°C).

8. Incubate the samples for at least 1 hour at room temperature to ensure complete actin polymerization before observation.

To prepare septin filaments in the absence of actin (Figure 6), we follow the same procedure as above, but replace the G-actin solution with G-buffer.

For experiments with preformed actin filaments at 1 μM, we first pre-polymerize actin at 24 μM (10% molar label ratio) in F-buffer for at least 1 hour at room temperature in the dark. We then follow the same procedure as above, but prepare the master buffer by taking into account that pre-polymerized F-actin contains 50 mM KCl and 2 mM $MgCl_2$.

[Insert Figure 5 here]

**Figure 5.** TIRF images of *in vitro* reconstituted actin-septin co-assembly, showing AF594-actin with 10% molar label ratio (left), AF488-Drosophila septins with 10% molar label ratio (middle) and the composite image (right). The concentrations of actin and Drosophila septins are 1 μM and 0.1 μM,

respectively. Under these conditions, the septins are predominantly present as hexamers (Mavrakis et al., 2014). Scale bar, 10 μm.

[Insert Figure 6 here]

**Figure 6.** TIRF images of AF488-Drosophila septin bundles at 1 μM (A) and of GFP-tagged Drosophila septin bundles at 1 μM (B). The spotty appearance of AF488-labeled bundles suggests that the effective labelling stoichiometry is below the nominal ratio of 10%. Scale bars, 10 μm.

## 7. TIRF microscopy

We image actin-septin filament assembly near the surface of the passivated coverslips by TIRF microscopy, which is ideally suited to provide a high signal-to-noise ratio for *in vitro* surface assays. Samples are imaged with a Nikon Apo TIRF × 100/1.49 NA oil objective mounted on an Eclipse Ti microscope (Nikon) using 491 nm and 561 nm laser lines and imaged with a QuantEM 512SC EMCCD camera (Photometrics). We generally use exposure times of 100-200 ms and optimize the laser power of the 488 nm and 561 nm laser lines to maximize the signal-to-noise ratio while minimizing photodamage (evident from the occurence of severing) of the actin and septin filaments.

**References**

Bertin, A., McMurray, M. A., Grob, P., Park, S. S., Garcia, G., 3rd, Patanwala, I., . . . Nogales, E. (2008). Saccharomyces cerevisiae septins: supramolecular organization of heterooligomers and the mechanism of filament assembly. *Proc Natl Acad Sci U S A, 105*(24), 8274-8279. doi:10.1073/pnas.0803330105

Booth, E. A., Vane, E. W., Dovala, D., & Thorner, J. (2015). A Forster Resonance Energy Transfer (FRET)-based System Provides Insight into the Ordered Assembly of Yeast Septin Hetero-octamers. *Journal of Biological Chemistry, 290*(47), 28388-28401. doi:10.1074/jbc.M115.683128

Bridges, A. A., Zhang, H., Mehta, S. B., Occhipinti, P., Tani, T., & Gladfelter, A. S. (2014). Septin assemblies form by diffusion-driven annealing on

membranes. *Proc Natl Acad Sci U S A, 111*(6), 2146-2151. doi:10.1073/pnas.1314138111

Cordes, T., Vogelsang, J., & Tinnefeld, P. (2009). On the mechanism of Trolox as antiblinking and antibleaching reagent. *J Am Chem Soc, 131*(14), 5018-5019. doi:10.1021/ja809117z

Diebold, M. L., Fribourg, S., Koch, M., Metzger, T., & Romier, C. (2011). Deciphering correct strategies for multiprotein complex assembly by co-expression: application to complexes as large as the histone octamer. *J Struct Biol, 175*(2), 178-188. Retrieved from http://www.ncbi.nlm.nih.gov/entrez/query.fcgi?cmd=Retrieve&db=PubMed&dopt=Citation&list_uids=21320604

Fares, H., Peifer, M., & Pringle, J. R. (1995). Localization and possible functions of Drosophila septins. *Mol Biol Cell, 6*(12), 1843-1859. Retrieved from http://www.ncbi.nlm.nih.gov/pubmed/8590810

Farkasovsky, M., Herter, P., Voss, B., & Wittinghofer, A. (2005). Nucleotide binding and filament assembly of recombinant yeast septin complexes. *Biol Chem, 386*(7), 643-656. doi:10.1515/BC.2005.075

Field, C. M., al-Awar, O., Rosenblatt, J., Wong, M. L., Alberts, B., & Mitchison, T. J. (1996). A purified Drosophila septin complex forms filaments and exhibits GTPase activity. *J Cell Biol, 133*(3), 605-616. Retrieved from http://www.ncbi.nlm.nih.gov/entrez/query.fcgi?cmd=Retrieve&db=PubMed&dopt=Citation&list_uids=8636235

Frazier, J. A., Wong, M. L., Longtine, M. S., Pringle, J. R., Mann, M., Mitchison, T. J., & Field, C. (1998). Polymerization of purified yeast septins: evidence that organized filament arrays may not be required for septin function. *J Cell Biol, 143*(3), 737-749. Retrieved from http://www.ncbi.nlm.nih.gov/pubmed/9813094

Garcia, G., 3rd, Bertin, A., Li, Z., Song, Y., McMurray, M. A., Thorner, J., & Nogales, E. (2011). Subunit-dependent modulation of septin assembly: budding yeast septin Shs1 promotes ring and gauze formation. *J Cell Biol, 195*(6), 993-1004. doi:10.1083/jcb.201107123

Green, M. R., & Sambrook, J. (2012). *Molecular Cloning: A Laboratory Manual (Fourth Edition)*: Cold Spring Harbor Laboratory Press.

Hartwell, L. H. (1971). Genetic control of the cell division cycle in yeast. IV. Genes controlling bud emergence and cytokinesis. *Exp Cell Res, 69*(2), 265-276. Retrieved from http://www.ncbi.nlm.nih.gov/pubmed/4950437

Hartwell, L. H., Culotti, J., Pringle, J. R., & Reid, B. J. (1974). Genetic control of the cell division cycle in yeast. *Science, 183*(4120), 46-51. Retrieved from http://www.ncbi.nlm.nih.gov/pubmed/4587263

Hsu, S. C., Hazuka, C. D., Roth, R., Foletti, D. L., Heuser, J., & Scheller, R. H. (1998). Subunit composition, protein interactions, and structures of the mammalian brain sec6/8 complex and septin filaments. *Neuron, 20*(6), 1111-1122. Retrieved from http://www.ncbi.nlm.nih.gov/pubmed/9655500

Huijbregts, R. P., Svitin, A., Stinnett, M. W., Renfrow, M. B., & Chesnokov, I. (2009). Drosophila Orc6 facilitates GTPase activity and filament formation of the septin complex. *Mol Biol Cell, 20*(1), 270-281. doi:10.1091/mbc.E08-07-0754

John, C. M., Hite, R. K., Weirich, C. S., Fitzgerald, D. J., Jawhari, H., Faty, M., . . . Steinmetz, M. O. (2007). The Caenorhabditis elegans septin complex is nonpolar. *EMBO J, 26*(14), 3296-3307. doi:10.1038/sj.emboj.7601775

Kim, M. S., Froese, C. D., Estey, M. P., & Trimble, W. S. (2011). SEPT9 occupies the terminal positions in septin octamers and mediates polymerization-dependent functions in abscission. *J Cell Biol, 195*(5), 815-826. doi:10.1083/jcb.201106131

Kim, M. S., Froese, C. D., Xie, H., & Trimble, W. S. (2012). Uncovering principles that control septin-septin interactions. *Journal of Biological Chemistry, 287*(36), 30406-30413. doi:10.1074/jbc.M112.387464

Kinoshita, M., Field, C. M., Coughlin, M. L., Straight, A. F., & Mitchison, T. J. (2002). Self- and actin-templated assembly of Mammalian septins. *Dev Cell, 3*(6), 791-802. Retrieved from http://www.ncbi.nlm.nih.gov/entrez/query.fcgi?cmd=Retrieve&db=PubMed&dopt=Citation&list_uids=12479805

Mavrakis, M., Azou-Gros, Y., Tsai, F. C., Alvarado, J., Bertin, A., Iv, F., . . . Lecuit, T. (2014). Septins promote F-actin ring formation by crosslinking actin filaments into curved bundles. *Nat Cell Biol, 16*(4), 322-334. doi:10.1038/ncb2921

Nakatsuru, S., Sudo, K., & Nakamura, Y. (1994). Molecular cloning of a novel human cDNA homologous to CDC10 in Saccharomyces cerevisiae. *Biochem Biophys Res Commun, 202*(1), 82-87. doi:10.1006/bbrc.1994.1896

Neufeld, T. P., & Rubin, G. M. (1994). The Drosophila peanut gene is required for cytokinesis and encodes a protein similar to yeast putative bud neck filament proteins. *Cell, 77*(3), 371-379. Retrieved from http://www.ncbi.nlm.nih.gov/entrez/query.fcgi?cmd=Retrieve&db=PubMed&dopt=Citation&list_uids=8181057

Nishihama, R., Onishi, M., & Pringle, J. R. (2011). New insights into the phylogenetic distribution and evolutionary origins of the septins. *Biol Chem, 392*(8-9), 681-687. doi:10.1515/BC.2011.086

Pan, F., Malmberg, R. L., & Momany, M. (2007). Analysis of septins across kingdoms reveals orthology and new motifs. *BMC Evol Biol, 7*, 103. doi:10.1186/1471-2148-7-103

Pardee, J. D., & Spudich, J. A. (1982). Purification of muscle actin. *Methods Cell Biol, 24*, 271-289. Retrieved from http://www.ncbi.nlm.nih.gov/pubmed/7098993

Popp, D., Yamamoto, A., Iwasa, M., & Maeda, Y. (2006). Direct visualization of actin nematic network formation and dynamics. *Biochem Biophys Res Commun, 351*(2), 348-353. doi:10.1016/j.bbrc.2006.10.041

Renz, C., Johnsson, N., & Gronemeyer, T. (2013). An efficient protocol for the purification and labeling of entire yeast septin rods from E.coli for quantitative in vitro experimentation. *BMC Biotechnol, 13*, 60. doi:10.1186/1472-6750-13-60

Sadian, Y., Gatsogiannis, C., Patasi, C., Hofnagel, O., Goody, R. S., Farkasovsky, M., & Raunser, S. (2013). The role of Cdc42 and Gic1 in the regulation of septin filament formation and dissociation. *Elife, 2*, e01085. doi:10.7554/eLife.01085

Sellin, M. E., Sandblad, L., Stenmark, S., & Gullberg, M. (2011). Deciphering the rules governing assembly order of mammalian septin complexes. *Mol Biol Cell, 22*(17), 3152-3164. doi:10.1091/mbc.E11-03-0253

Serrao, V. H., Alessandro, F., Caldas, V. E., Marcal, R. L., Pereira, H. D., Thiemann, O. H., & Garratt, R. C. (2011). Promiscuous interactions of human septins:

the GTP binding domain of SEPT7 forms filaments within the crystal. *FEBS Lett, 585*(24), 3868-3873. doi:10.1016/j.febslet.2011.10.043

Sheffield, P. J., Oliver, C. J., Kremer, B. E., Sheng, S., Shao, Z., & Macara, I. G. (2003). Borg/septin interactions and the assembly of mammalian septin heterodimers, trimers, and filaments. *Journal of Biological Chemistry, 278*(5), 3483-3488. doi:10.1074/jbc.M209701200

Shi, X., Lim, J., & Ha, T. (2010). Acidification of the oxygen scavenging system in single-molecule fluorescence studies: in situ sensing with a ratiometric dual-emission probe. *Anal Chem, 82*(14), 6132-6138. doi:10.1021/ac1008749

Sirajuddin, M., Farkasovsky, M., Hauer, F., Kuhlmann, D., Macara, I. G., Weyand, M., . . . Wittinghofer, A. (2007). Structural insight into filament formation by mammalian septins. *Nature, 449*(7160), 311-315. doi:10.1038/nature06052

Sirajuddin, M., Farkasovsky, M., Zent, E., & Wittinghofer, A. (2009). GTP-induced conformational changes in septins and implications for function. *Proc Natl Acad Sci U S A, 106*(39), 16592-16597. doi:10.1073/pnas.0902858106

Soares e Silva, M., Depken, M., Stuhrmann, B., Korsten, M., MacKintosh, F. C., & Koenderink, G. H. (2011). Active multistage coarsening of actin networks driven by myosin motors. *Proc Natl Acad Sci U S A, 108*(23), 9408-9413. doi:10.1073/pnas.1016616108

Versele, M., Gullbrand, B., Shulewitz, M. J., Cid, V. J., Bahmanyar, S., Chen, R. E., . . . Thorner, J. (2004). Protein-protein interactions governing septin heteropentamer assembly and septin filament organization in Saccharomyces cerevisiae. *Mol Biol Cell, 15*(10), 4568-4583. doi:10.1091/mbc.E04-04-0330

Zent, E., Vetter, I., & Wittinghofer, A. (2011). Structural and biochemical properties of Sept7, a unique septin required for filament formation. *Biol Chem, 392*(8-9), 791-797. doi:10.1515/BC.2011.082

Zent, E., & Wittinghofer, A. (2014). Human septin isoforms and the GDP-GTP cycle. *Biol Chem, 395*(2), 169-180. doi:10.1515/hsz-2013-0268

**Figure legends**

**Figure 1**. Overview of cloning strategy for co-expressing recombinant septins using the pnEA-vH and pnCS vectors of the pET-MCN series (Diebold et al., 2011). Here this strategy is used for the generation of recombinant Drosophila (DSep1-DSep2-Pnut) and human (hSep2-hSep6-hSep7) septin hexamers. See text for details.

**Figure 2.** Schematic overview of the two-tag purification scheme for isolating stoichiometric human septin hexamers (the same applies to Drosophila septins by analogy). See text for details. Understanding the principles of septin complex assembly is an intense topic of investigation, and the presence and stability of intermediate complexes has not been fully documented. We show selected populations of human septin complexes in the cell lysate (monomers, homo- and heterodimers,

heterotrimers and hexamers) based on the isolation and characterization of such recombinant and native complexes (Kim et al., 2012; Mavrakis et al., 2014; Sellin et al., 2011; Serrao et al., 2011; Sheffield et al., 2003; Sirajuddin et al., 2007; Sirajuddin et al., 2009; Zent et al., 2011; Zent & Wittinghofer, 2014).

**Figure 3.** Characterization of purified septin complexes by SDS-PAGE (unlabeled and GFP-labeled Drosophila septin hexamers, left and right panels in A, respectively) and by 2D single particle analysis of electron microscopy images (B, see chapter by Aurelie Bertin).

**Figure 4.** A flow cell assembled by melting strips of Parafilm between a coverslip and a glass slide.

**Figure 5.** TIRF images of *in vitro* reconstituted actin-septin co-assembly. The concentrations of actin and Drosophila septins are 1 μM and 0.1 μM, respectively. Scale bar, 10 μm.

**Figure 6.** TIRF images of AF488-Drosophila septin bundles at 1 μM (A) and of GFP-tagged Drosophila septin bundles at 1 μM (B). Scale bars, 10 μm.

**Acknowledgments**

We thank M. Kuit-Vinkenoog for G-actin purification and F. Iv for septin purification. The research leading to these results has received funding from CNRS, from two PHC Van Gogh grants (no. 25005UA and no. 28879SJ, ministères des Affaires étrangères et de l'Enseignement supérieur et de la Recherche), and from the European Research Council under the European Union's Seventh Framework Programme (FP/2007-2013) / ERC Grant Agreement n. [335672].

## septin co-expression using the pET-MCN vector series

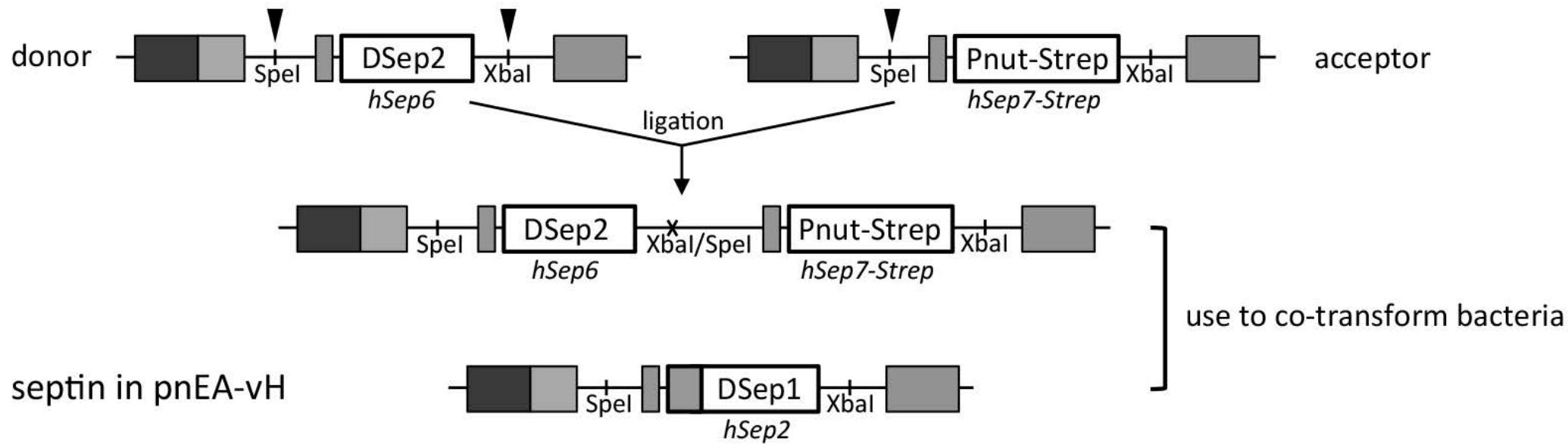

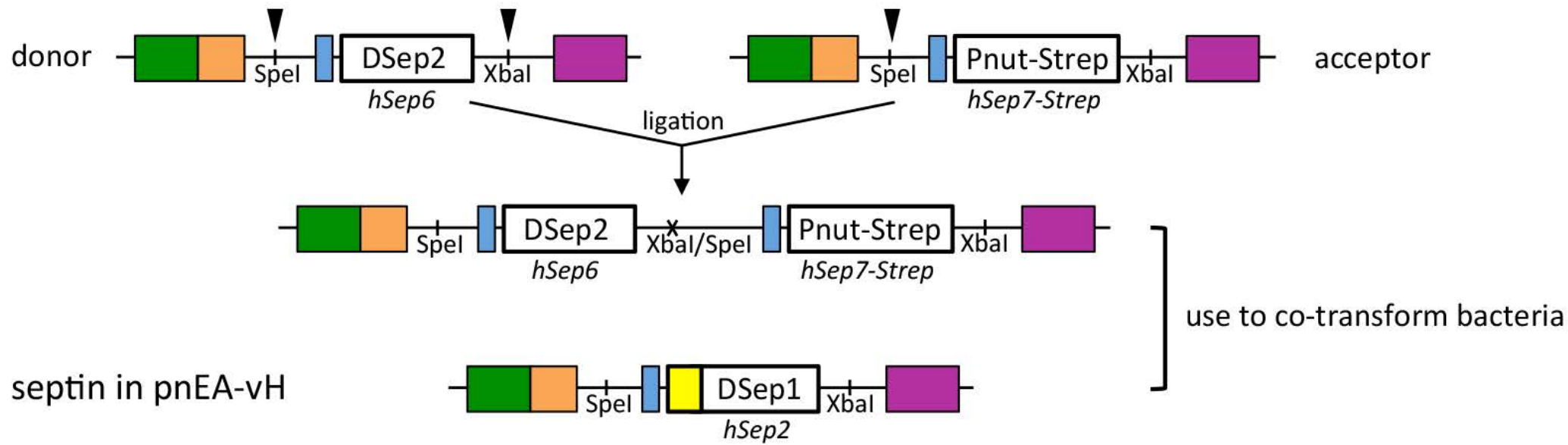

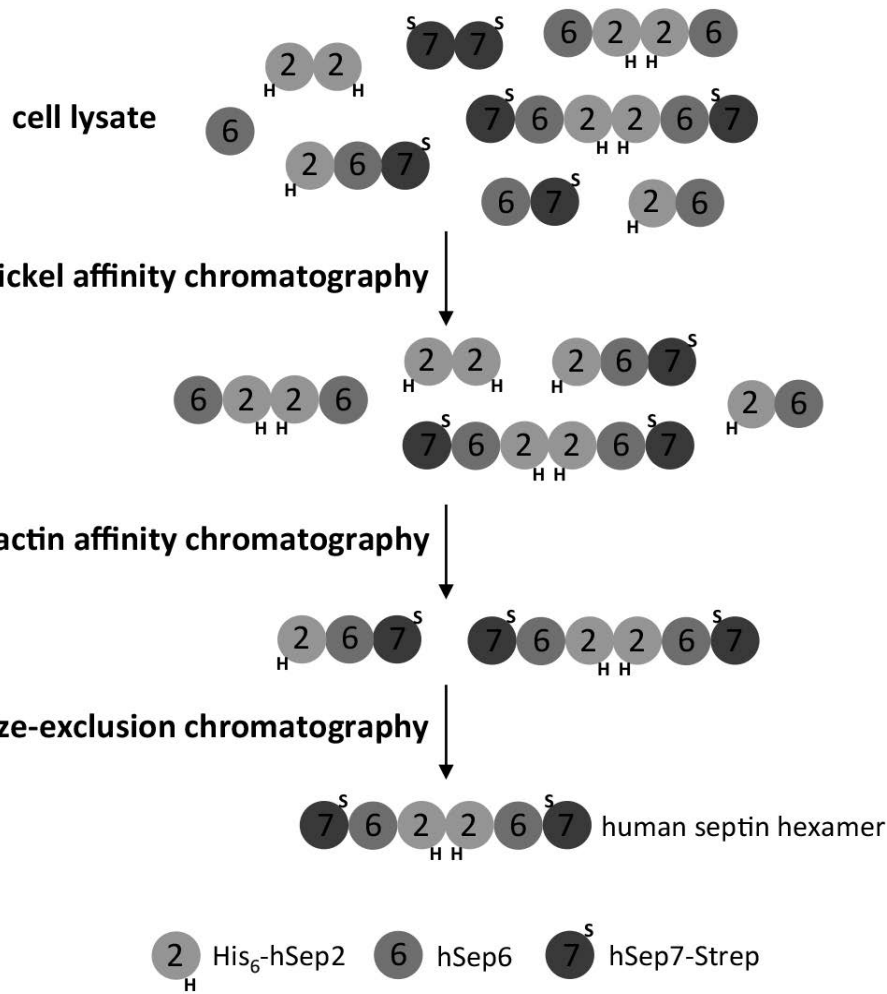

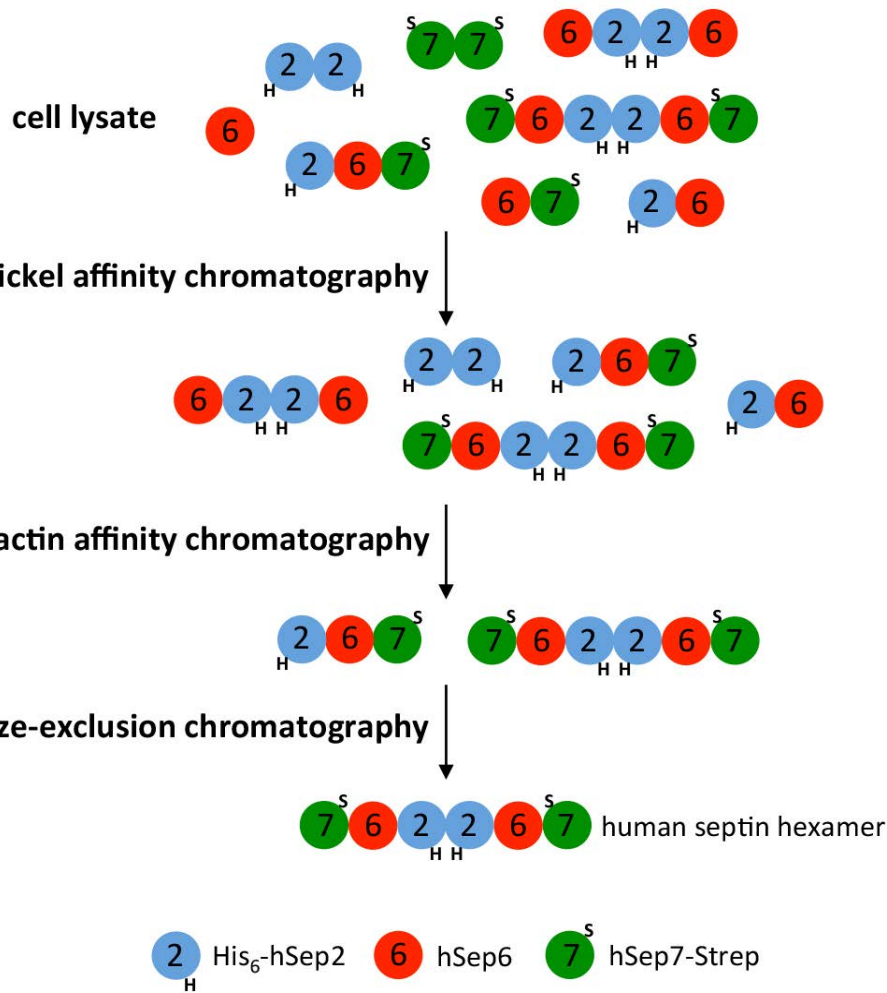

**A** SDS-PAGE

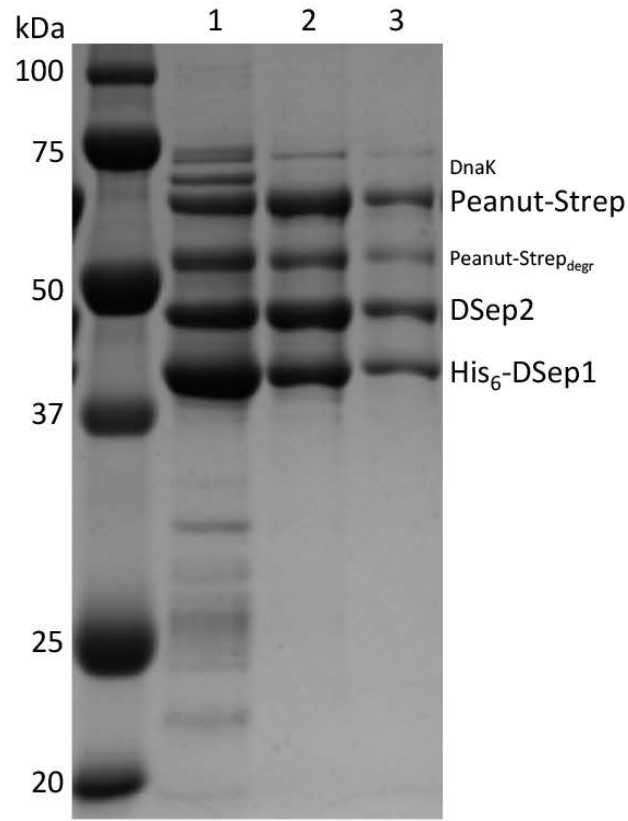

**B** 2D analysis by EM

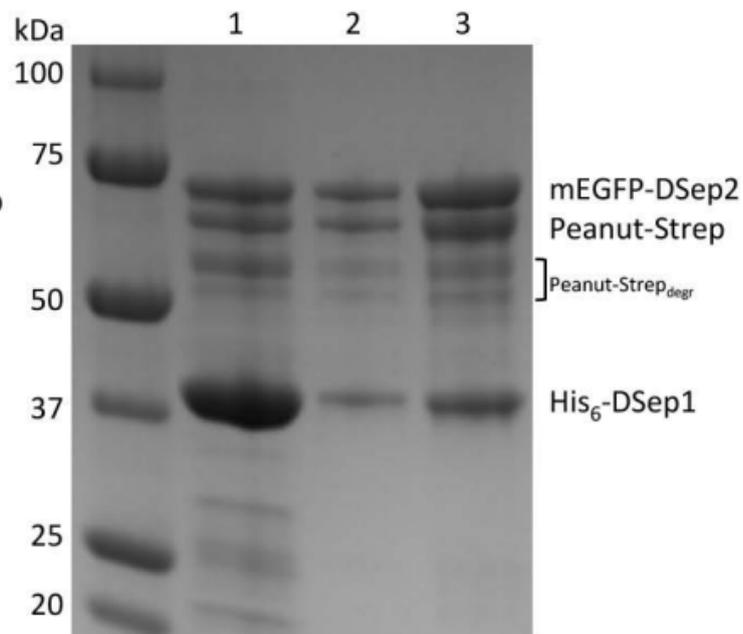

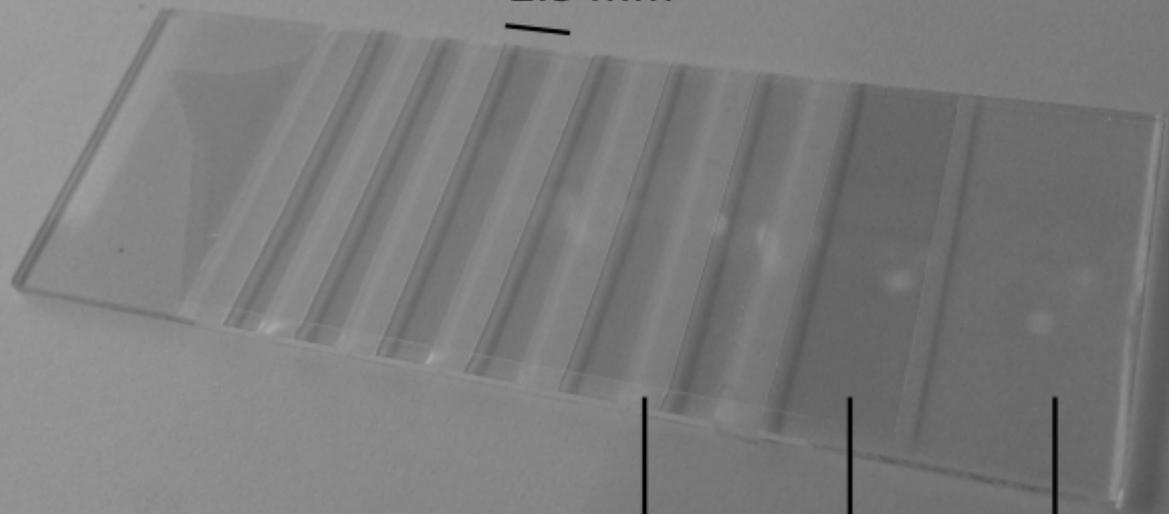

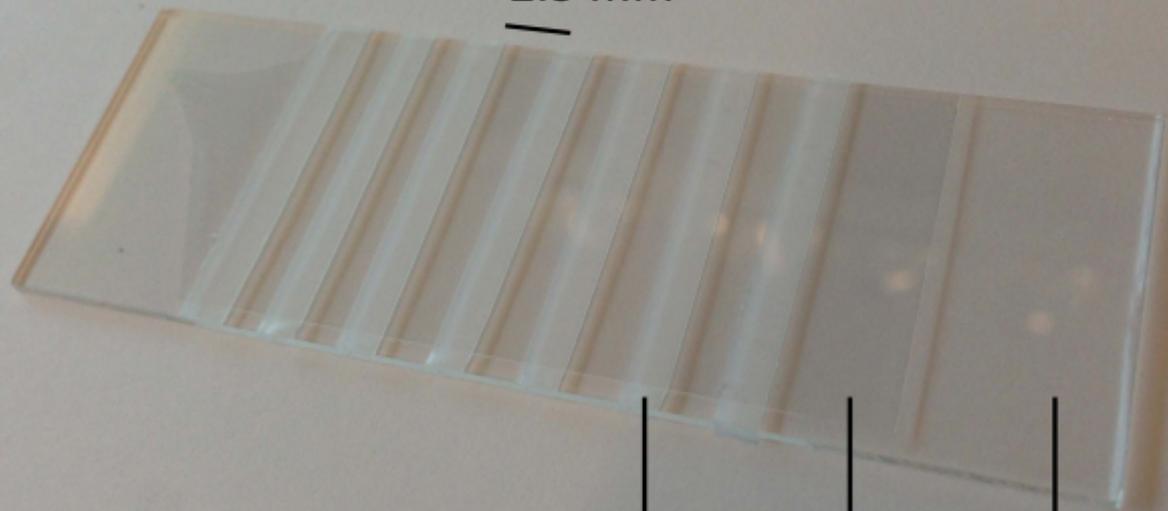

~ 2.5 mm

parafilm   coverslip   glass slide

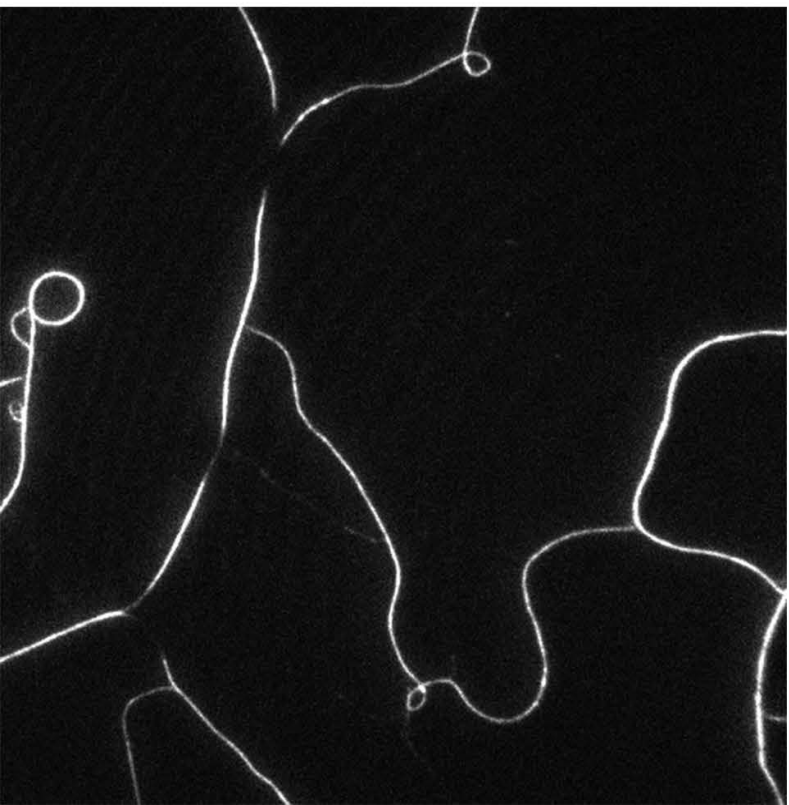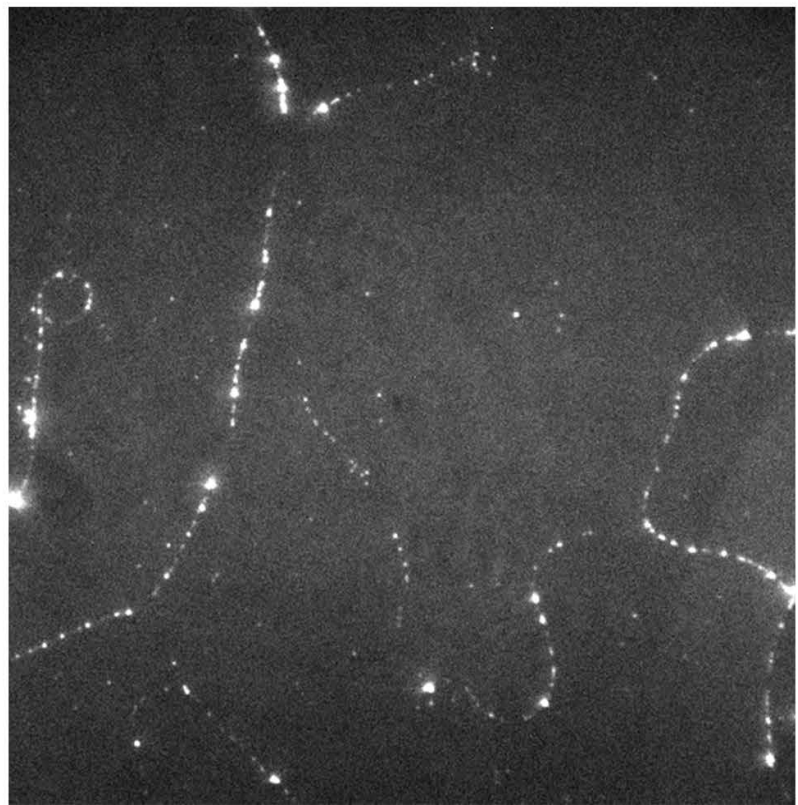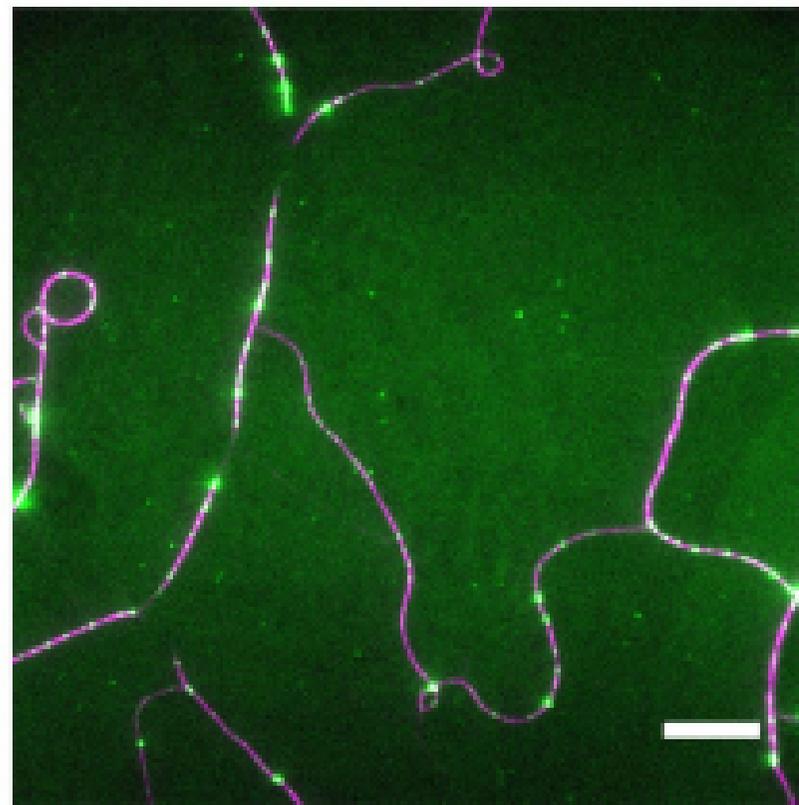

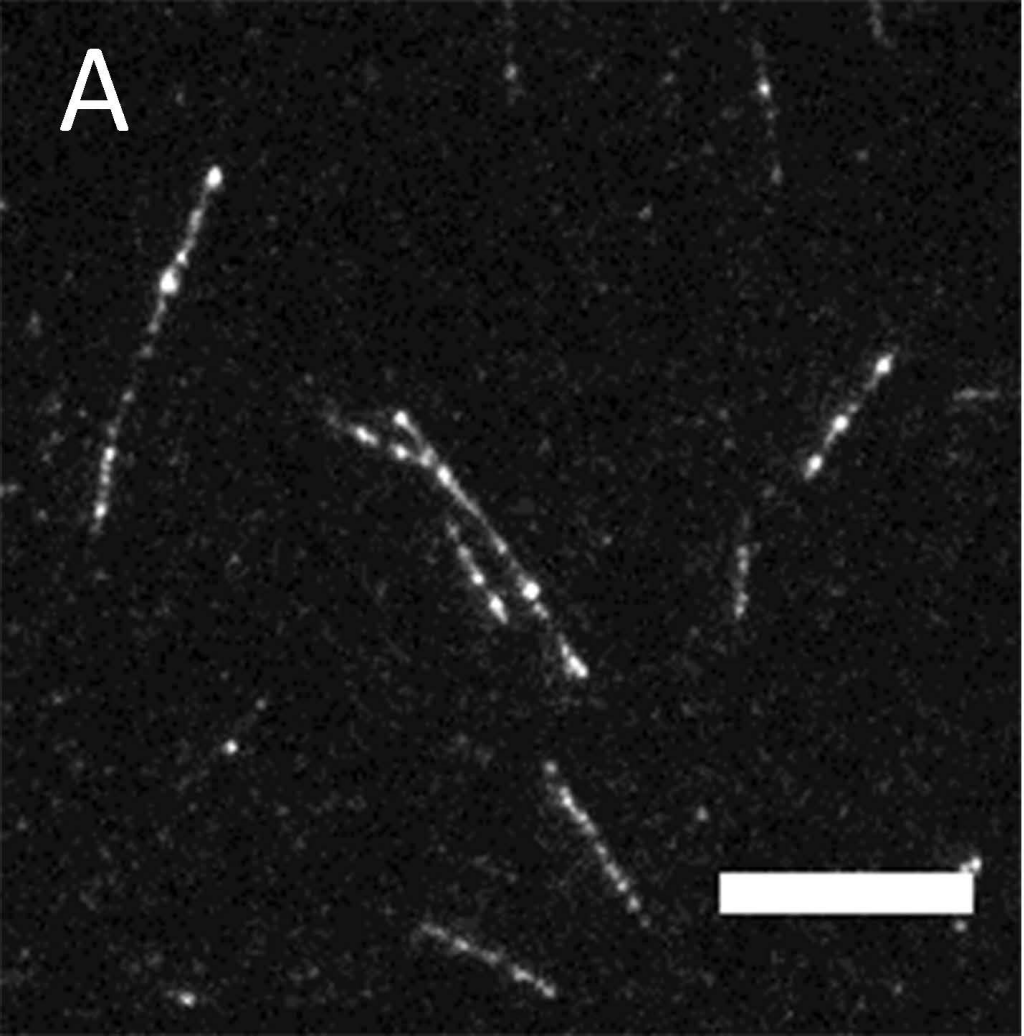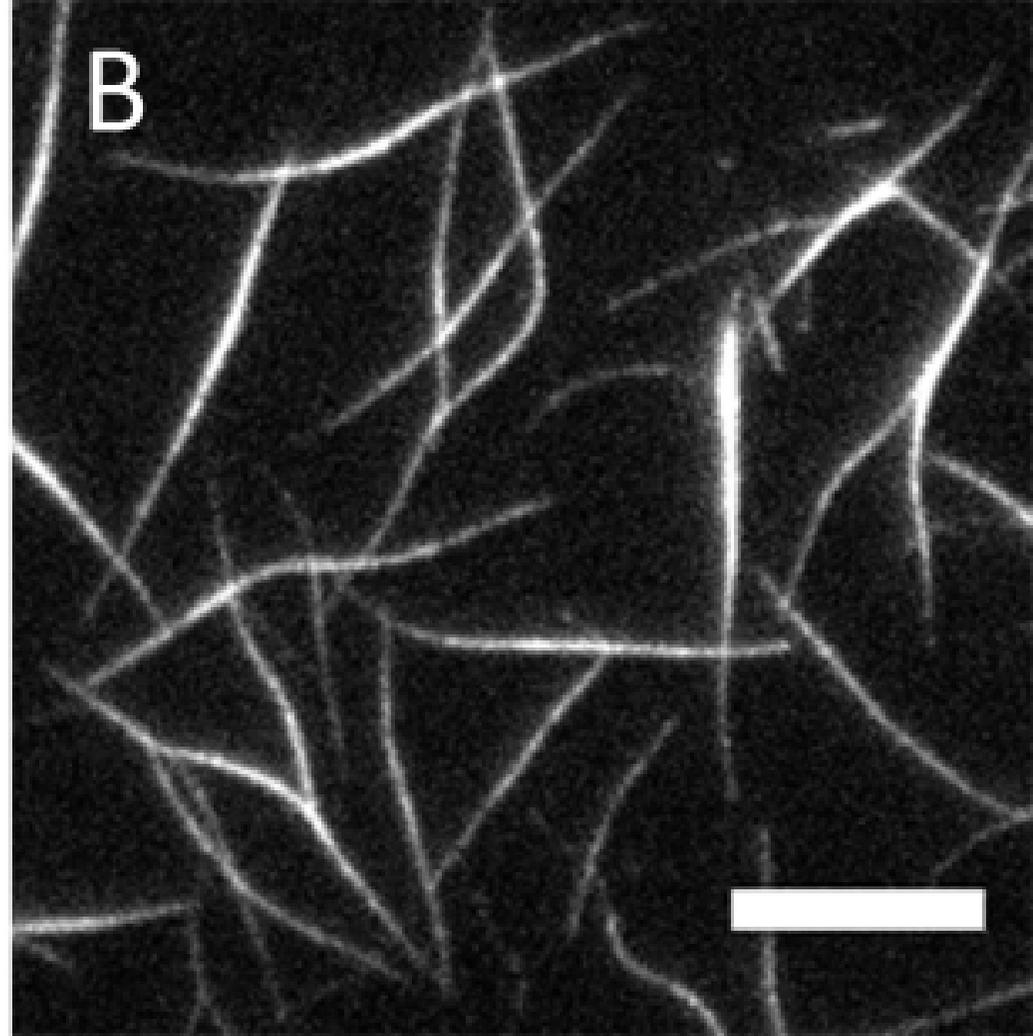